\documentclass[preprint]{aastex}
\usepackage[utf8]{inputenc}
\usepackage{textgreek}
\usepackage{tablefootnote}

\usepackage{natbib}
\newcommand{\del}{\textdelta \hspace{1 mm}}
\newcommand{\um}{\textmu m}
\newcommand{\msun}{M$_{\odot}$}
\newcommand{\rsun}{R$_{\odot}$}

\begin{document}
\title{Resolving the delta Andromedae spectroscopic binary with direct imaging}
\author{Michael Bottom\altaffilmark{1}, Jonas Kuhn\altaffilmark{2, 3}, Bertrand Mennesson\altaffilmark{2},  Dimitri Mawet\altaffilmark{1}, Jean C. Shelton\altaffilmark{2}, J. Kent Wallace\altaffilmark{2}, Eugene Serabyn\altaffilmark{2}}
\date{March 2015}
\email{mbottom@caltech.edu}
\altaffiltext{1}{Cahill Center for Astronomy and Astrophysics , California Institute of Technology, MC 249-17, Pasadena, CA 91125, USA}
\altaffiltext{2}{Jet Propulsion Laboratory, California Institute of Technology, Pasadena, CA, 91109 USA}
\altaffiltext{3}{(current address) Institute for Astronomy, ETH Zurich, Wolfgang-Pauli-Strasse 27, CH-8093 Zurich, Switzerland}

\begin{abstract}
We present a direct image of the innermost companion to the red giant \del Andromedae using the Stellar Double Coronagraph at the Palomar Observatory.  We use a Markov-chain Monte Carlo based algorithm to simultaneously reduce the data and perform astrometry and photometry of the companion.  {We determine that the companion is most likely a main-sequence K-type star and is certainly not the previously hypothesized white dwarf.}
\end{abstract}
\maketitle

\section{Introduction}
 \del Andromedae (K3 III) is red giant with a visual magnitude of 3.28.  It has a UV excess which implies a hot, high-velocity wind and a 60 and 100 \textmu m excess \citep{1987MNRAS.224...93J} which is most likely due to a debris disk \citep{2003ApJ...598..636D}.  It is the brightest star in a quadruple system; of the outer companions (28.7 and 48 arc seconds), the first has been classified as an M2 V star with $V$ = 11.3, probably physically associated with the primary as it shares the same proper motion \citep{1976JRASC..70...23B}.  The outer component  does not share the proper motion of the system and is most likely a background object (ibid).
 
{\del Andromedae is a spectroscopic binary with a rather long period of about 57 years \citep{2008AJ....135..209M}; see Table \ref{delandprops} for a summary of its physical properties.  The presence of the secondary has been confirmed photocentrically and astrometrically \citep{2002A&A...391..647G}, with both sources deriving an eccentricity of about 0.5.  }The companion has been conjectured to be either a main-sequence star later than G-type \citep{1987MNRAS.224...93J}  or white dwarf near the Chandrasekhar limit \citep{2002A&A...391..647G}.  It has never been directly imaged, however, due to the secondary's faintness and proximity to the primary.  In this work we image the companion for the first time and measure its magnitude and separation. {We find that the secondary is at approximately the expected separation and determine that it is not a white dwarf, but most likely a main sequence star of K-type}.  This work demonstrates the potential of high contrast imaging with low inner working angles applied to spectroscopic binaries.

\begin{table}[h]
\centering 
\begin{tabular}{lll}
\multicolumn{3}{c}{Physical properties of $\delta$ And} \\
\hline\\
Mass (M$_A$+M$_B$)            & 2.6 $\pm$ 0.4 M$_{\odot}$   & {\citet{2002A&A...391..647G}}\\
a$_A$+a$_B$                   & 0.62 $\pm$ 0.04''           & {\citet{2002A&A...391..647G}}\\
{Period}			& {57.6 $\pm$  1.09 yr} & {{\citet{2008AJ....135..209M}}}\\
Radius                        & 13.6 $\pm$ 0.3 R$_{\odot}$  & {\citet{2011A&A...526A.100P}}\\
Luminosity                    & 68 $\pm$ 4 L$_{\odot}$      & {\citet{2011A&A...526A.100P}}\\
Surface gravity (log g)       & 2.0 $\pm$ 0.3               & {\citet{1987MNRAS.224...93J}}\\
Temperature                   & 4315 $\pm$ 9 K              & {\citet{2008AJ....135..209M}}\\
Metallicity {[}Fe/H{]}        & 0.04 dex                    & {\citet{2008AJ....135..209M}}\\
Rotational velocity ($v$ sin $i$) & 6.5 km/s                & {\citet{2008AJ....135..209M}}\\
Age                           & 3.2 Gyr                     & {\citet{2003ApJ...598..636D} (v. uncertain)} \\
{Parallax}		      & {0.032 $\pm$ 0.001'' }& {{\citet{2007A&A...474..653V}}}\\
\hline\\
\del And b properties (this work) & {} &{}\\
\hline
$\Delta$M (Bracket-\textgamma)    & 6.22 $\pm$ 0.05        & {{$\lambda_c$ = 2.18 $\mu$m, $\Delta \lambda$ =0.03 $\mu$m}} \\  
Angular Separation                          & 0.357\  $\pm$ 0.0035 ''  & { } \\
Position angle                              & 56 $\pm$ 1$^{\circ}$  & {} \\
Physical Separation                         & 11.55\ $\pm$ 0.13 AU & {parallax from \textit{Hipparcos}, as above}\\
Spectral type                               & K4 $\pm$ 2            & {derived from ATLAS9 spectra} \\
\end{tabular}
\caption{Previously measured properties of \del And and newly measured properties of the companion}
\label{delandprops}
\end{table}
 
\section{Instrumentation, Observations and Data Analysis}
\subsection{Instrumentation}
The Stellar Double Coronagraph (SDC) is a JPL-developed instrument designed for high-contrast imaging of close-in companions to stars, particularly exoplanets.  It uses two optical vortices in series to simultaneously diffract starlight out of the pupil of the instrument and partially correct for the secondary obscuration of the telescope \citep{2011OptL...36.1506M}.  It has an inner working angle of approximately 1\textlambda/D, or 90 mas in K-band (2.2 \um) behind the 5 m Hale telescope.  It is installed between the P3K adaptive optics system \citep{2013ApJ...776..130D} and the near-IR imager PHARO \citep{2001PASP..113..105H}.

\subsection{Observations}
We observed \del And on October 8-9 2014 (UTC), during the course of normal science operations.  The seeing was 1.2'', with the adaptive optics system delivering a Strehl ratio of about 85\% at an airmass of 1.02-1.03.  Our observing strategy involved frequently dithering between the target star and a reference star, then using post-processing to subtract the speckle pattern from the target images using the reference (see the following section for more detail). A Bracket-\textgamma \ filter was used concurrently with neutral density filters to reduce the flux from the target when off the coronagraph to below detector saturation.  Absolute transmissivity of the neutral density filters was measured separately, and found to be consistent with \citet{2004ApJ...617.1330M}. Sky backgrounds were interspersed with the target and reference star observations; sky flats were taken five days later.  A summary of the observations is presented in Table \ref{obsprops}.

\begin{table}[ht]
\centering 
\begin{tabular}{lllll}
\multicolumn{4}{c}{Observing date: Oct. 9-10 2014, JD 2456939-10}\\
\hline\\
Target & Images & Filters &Exposure Time[s] & Purpose \\
\hline\\

\del And            & 29          & Br-\textgamma, ND2  & 9.91  &Photometry, Astrometry\\
\textbeta \  And    & 100         & Br-\textgamma, ND2  & 2.83  &Photometry\\
\del And            & 10          & Br-\textgamma, ND3  &2.83   &Non-coronagraphic, Photometry\\
\end{tabular}
\label{obsprops}
\caption{Summary of observations.}
\end{table}

\subsubsection{Data Analysis}
One of the main challenges in high contrast imaging is trying to remove speckles due to aberrations in the optics after the wavefront sensor.  There are a number of ways to tackle this contrast-limiting/quasi-static aberration problem; our strategy is sometimes called ``reference differential imaging'' \citep{2011ApJ...738L..12M}, which involves dithering between the target and a nearby star of similar visible magnitude, spectral type, and airmass.  This leads to a similar AO correction and gravity vector, ensuring a similar speckle pattern.  It is then possible to remove some of the speckles by either subtracting the reference image or using a more advanced image processing technique such as the Karhunen-Loeve eigenimage decomposition \citep{2012ApJ...755L..28S}. The latter method gives better results than the former in terms of contrast, but has the unfortunate side effect of reducing the flux of any nearby companions that might be in the image, therefore rendering accurate photometry difficult.  In this paper, we use a slightly different approach where we forward model the target image as a combination of a scaled reference image and a shifted, attenuated point-spread function image.  This method has some advantages that will be explained below.

We acquired coronagraphic images of \del Andromedae and the reference star, \textbeta \ \ Andromedae.  We aligned and median combined these images after flat-fielding, background subtraction and bad pixel removal. We derive a relative magnitude and offset between the star and companion using a Markov Chain Monte Carlo (MCMC) fitting algorithm \citep{2013PASP..125..306F}.  This is somewhat different than the usual approach to analyzing fluxes and positions, where one prioritizes maximising the signal to noise ratio of the companion, often performing astrometry and photometry separately.  Here the image reduction, raw photometry, and astrometry are all performed at the same time by the MCMC algorithm.  There are a few advantages to doing everything at once with MCMC.  First, one can measure the precision of the reduction algorithm much more accurately: the per-pixel uncertainties are Poissonian and straightforward to propagate in the model above.  Furthermore, the MCMC returns marginal likelihoods, which shows the precision in each parameter as well as any correlations.  Finally, one does not need an analytic model of the PSF but can use images of the instrumental point spread function taken off the coronagraph.  This reduces the number of parameters in the model, decreases degeneracy, and improves accuracy. 

\begin{figure*}[tbp]
\centering
\includegraphics[width=0.45\textwidth]{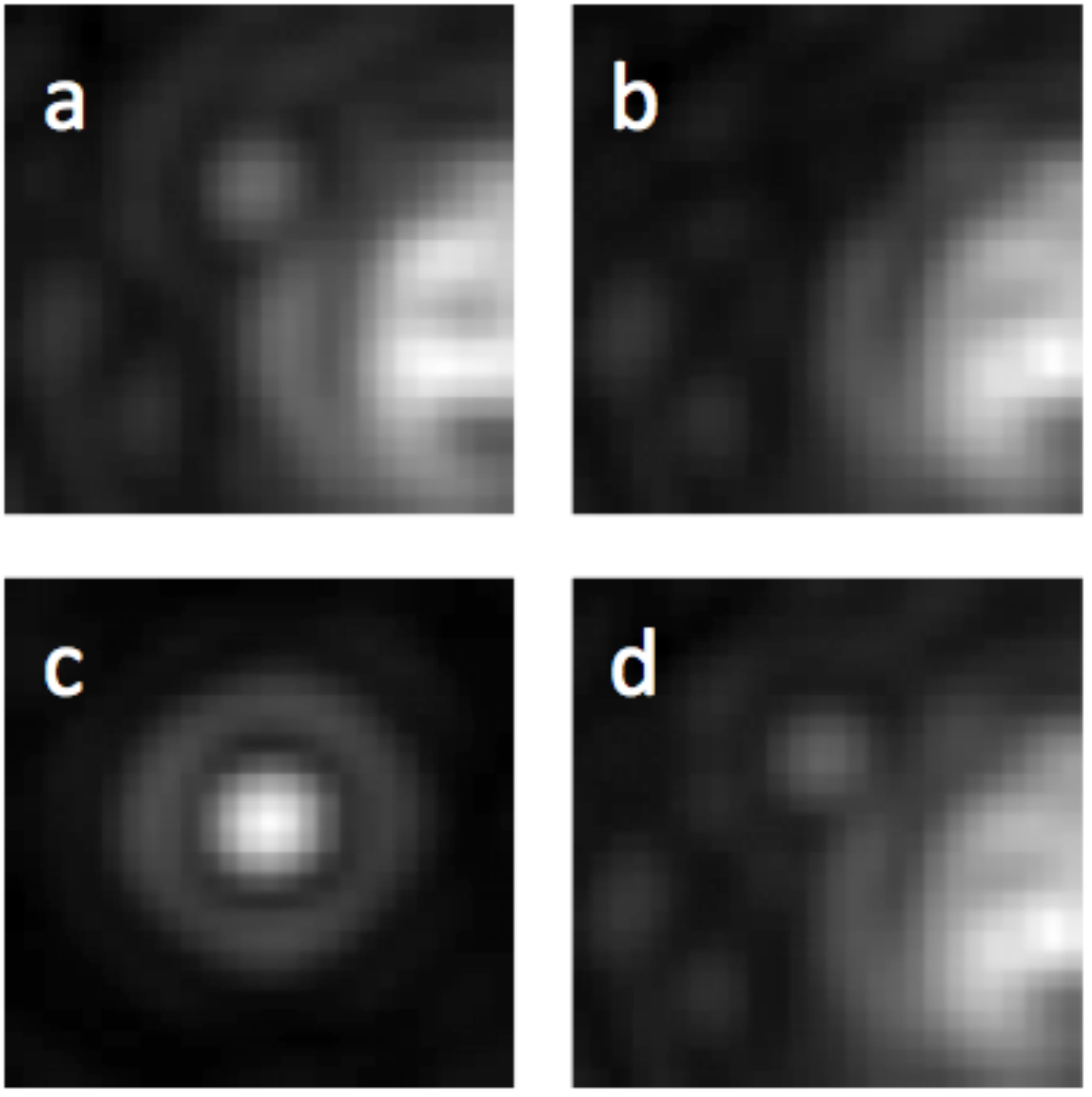}
\includegraphics[width=0.45\textwidth]{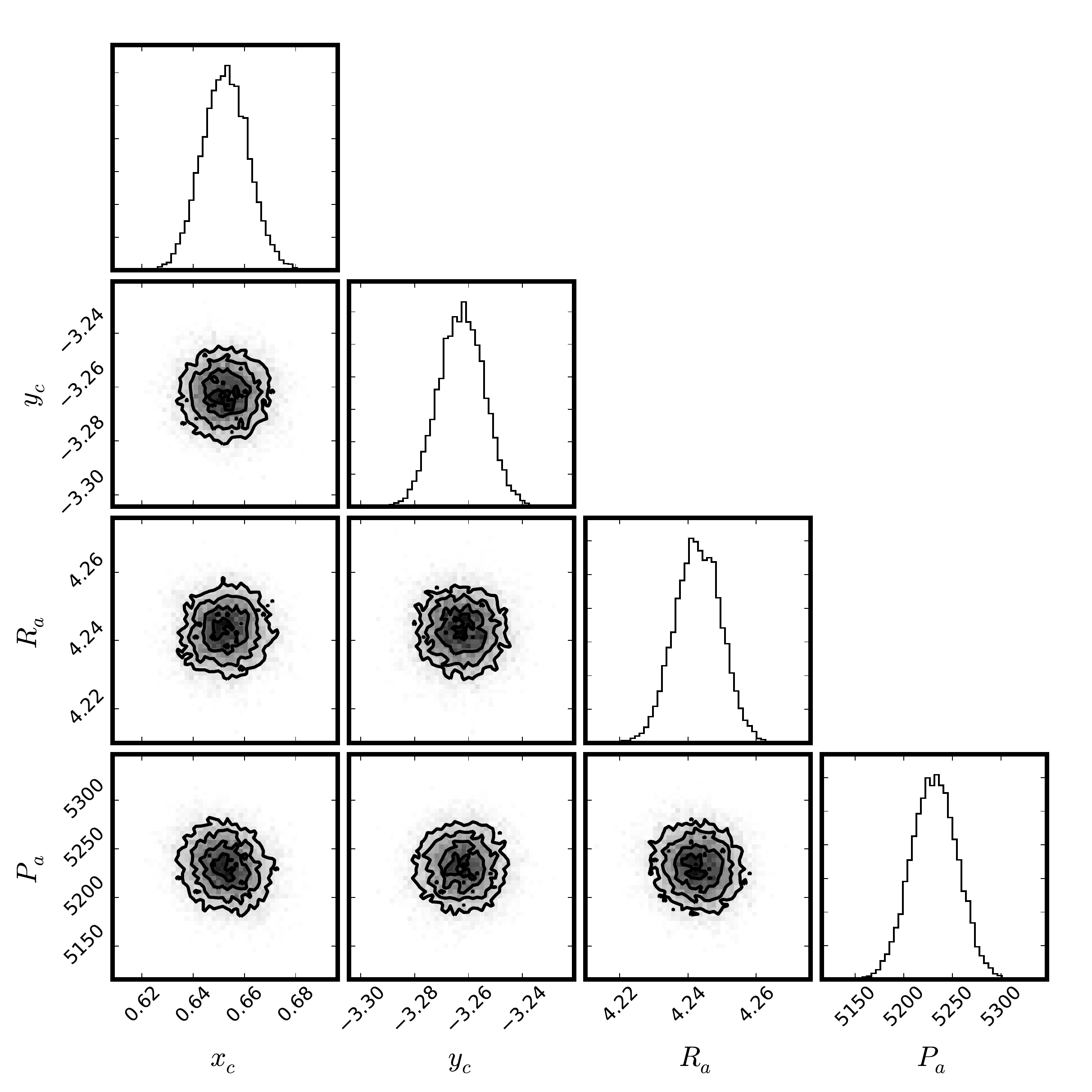}
\caption{Left: a) the background-subtracted target median image, b) the background-subtracted reference star median image, c) the  background-subtracted point-spread function image, d) best-fitting model from the MCMC algorithm combining images b) and c) attempting to match a) as explained above.  The stretch is nonlinear to better show the companion and speckles.  Right:  All the one and two dimensional projections of the posterior probability distributions of the pixel shifts ($x_c$, $y_c$, the reference background scaling factor ($R_a$, and the PSF amplitude used to fit the companion $P_a$.  The two-dimensional projections show very little covariance among any two parameters, and the marginal distribution histograms (along the diagonal) are nicely peaked.}
\label{mcmcout}
\end{figure*}

\begin{figure}
\epsscale{1}
\plotone{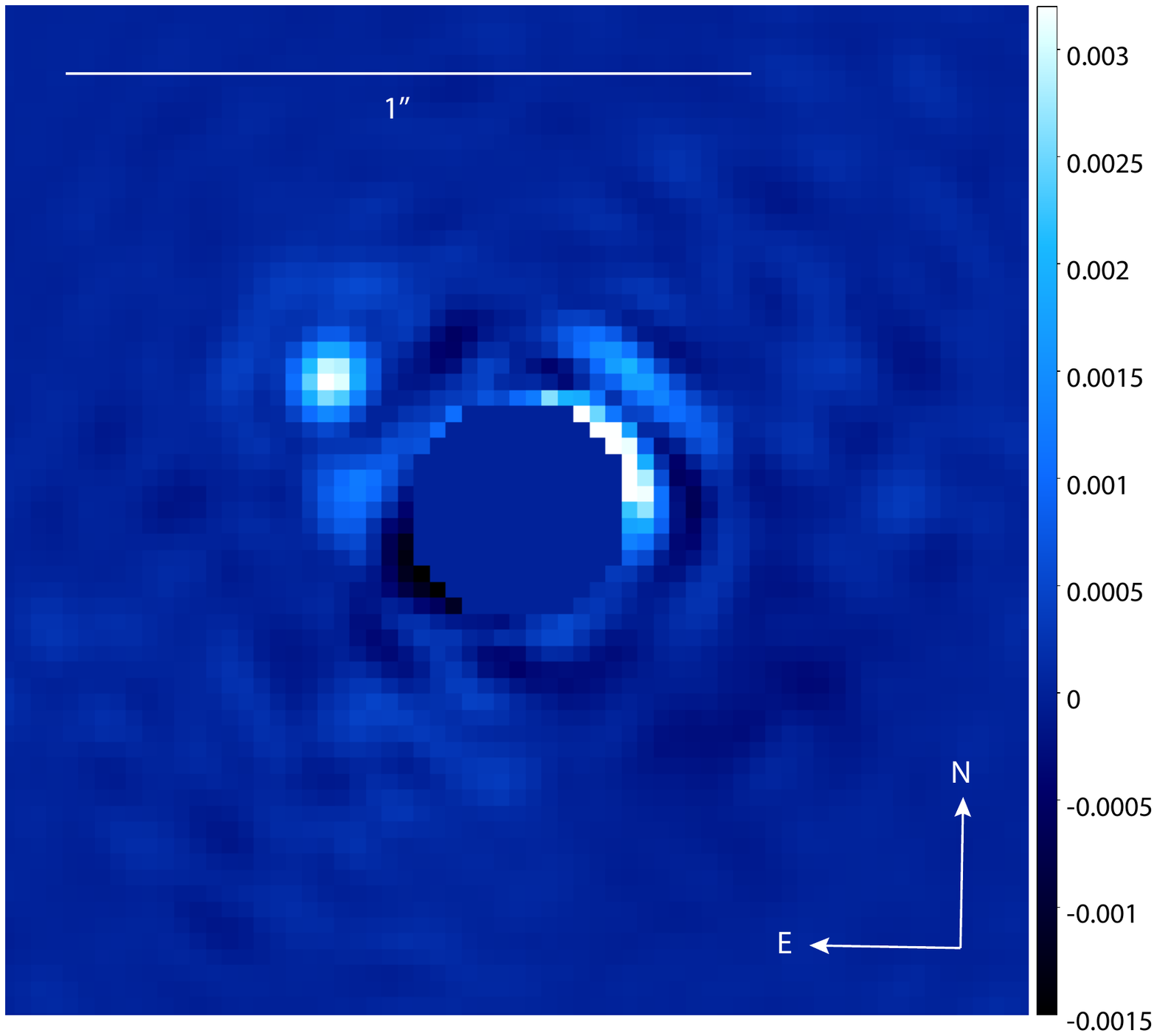}
\caption{The reduced, background-removed coronagraphic image of \del Andromedae.  The first Airy ring is visible around the companion.  The stretch in the image is linear.  The colorbar shows the relative intensity (as a fraction) compared to the primary}
\label{delandimage}
\end{figure}

The generative model for the image data is 

\begin{equation}
T[x,y] = R_a \cdot R[x,y] + P_a\cdot P[x-x_c, y-y_c]
\end{equation}

\noindent
where $T$ is the coronagraphic image of $\delta$ And, $R_a$ is a constant scale factor, $R$ is the coronagraphic reference image of $\beta$ And, $P_a$ is another constant scale factor, and $P$ is the point-spread function (ie, a unit-intensity normalized, non-coronagraphic image of a point source).  Images $T$, $R$, and $P$ are all single median images.  The indices $x$, $y$ refer to pixel coordinates, and the factors $x_c, y_c$ are shifts in point-spread function imaging data (ie, P[x-1, y-0.34] corresponds to a pixel shift of 1, 0.34).  The constant $R_a$ is to correct for the fact that the background speckle field in the science image is of a  different mean intensity, due to differing stellar magnitudes.  $P_a$ is the intensity scaling prefactor of the point spread function of the companion.  MCMC is used to solve  for $x_c$, $y_c$, $P_a$, and $R_a$ simultaneously; the results are presented in Figure \ref{mcmcout}.  The ``reference subtracted'' image, $T[x,y] - R_a\cdot R[x,y]$ is shown in Figure \ref{delandimage}.

In order to determine the relative brightness, we similarly use the unit-intensity PSF model to fit a non-coronagraphic image of \del And, and the derived intensity allows us to establish a relative intensity in magnitudes.  The uncertainty in relative intensity is dominated by the uncertainty on the neutral density filters' absolute extinctions.  For the companion location, the typical error in this case for $x_c$ and $y_c$ was about 0.01 pixels, or less than a milliarcsecond at 0.025''/pixel.  However, this is not the true uncertainty in \textit{separation} because the primary star's image is suppressed and distorted by the coronagraph and its true position is not obvious to calculate.  In order to locate the position of the primary, we imposed a waffle pattern on the deformable mirror of the adaptive optics during observations.  The waffle generates astrometric spots 3.9'' away from the primary, which can be used to locate the position of the star, and we verified our result using the Radon transform \citep{2014arXiv1409.6388P}.  The waffle centration has an uncertainty of about 0.1 pixels, which dominates the total separation uncertainty.

\section{Results and Conclusions}
The results of the above analysis are shown in Table \ref{delandprops}.  Judge et al. suggest that the secondary companion is either  a main sequence star later than G type or a white dwarf.  Gontcharov and Kiyaeva measure a mass fraction  $m_B/(m_A+m_B)$ = 0.5 $\pm$ 0.1 for the binary system and then favor the white dwarf assumption, suggesting an even split of mass between primary and secondary, placing the white dwarf very near the Chandrasekhar mass limit.

\begin{figure}[h]
\epsscale{1}
\plotone{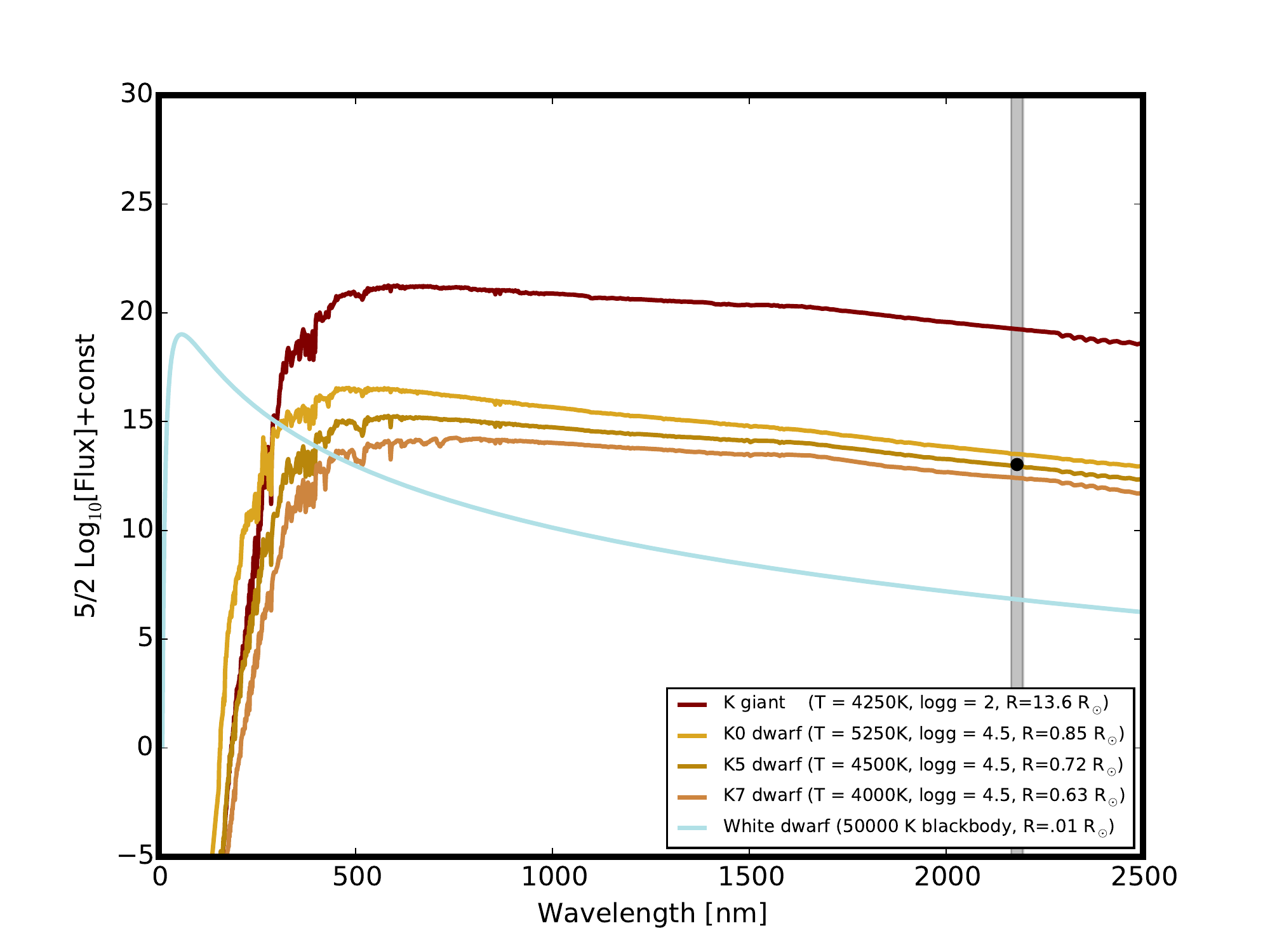}
\caption{Comparison of the approximate fluxes of \del And A, three K dwarfs, and a 50000K white dwarf.  The spectral models are from \citet{2004astro.ph..5087C}.  The width of the shaded bar is the span of the Bracket-\textgamma \  filter.  }
\label{comparison}
\end{figure}

However, assuming that the companion is a white dwarf, its radius is constrained to be about 0.01 \rsun, as white dwarfs of 0.5-1.4 \msun \ span the radius range of 0.014 to 0.005 \rsun.  Comparing the expected flux levels of a hot white dwarf to that of the primary (see Figure \ref{comparison}), the magnitude difference through the Bracket-\textgamma \  filter would be about 12 magnitudes, not the measured 6, a discrepancy of greater than 100 times our photometric uncertainty.  Furthermore, such a hot white dwarf would have a UV continuum that was not detected in Judge et al., who constrain the white dwarf's temperature to less than 10000K if it exists.  This low temperature would make the magnitude difference even more extreme.  The white dwarf possibility is thus definitively excluded.

On the other hand, the measured flux is consistent with a main-sequence K-type dwarf.  Shown in Figure \ref{comparison} are spectral models of K0, K5, and K7 stars, with our measured flux shown as a black dot.  For the \del And primary, the effective temperature and radius is taken from published results (Table \ref{delandprops}; note that the radius is known accurately from interferometry \citep{2011A&A...526A.100P}).  For the secondary, the effective temperatures and surface gravities are taken from \citet{2004astro.ph..5087C} and the radii are taken from \citet{2000asqu.book.....C}. While the formal photometric error is smaller than the size of the datapoint, lack of knowledge about the companion's radius and temperature make it impossible to give a completely specific spectral classification; the best we can say is that the companion is most likely a K-type dwarf.  Making a more accurate measurement of the secondary spectral type is possible in principle.  The simplest way would likely be a similarly precise coronagraphic measurement in $J$ band, as the $J$-$K$ colors of K dwarfs change by about 200 millimags over the spectral type.  Alternatively, an AO-fed integral field spectrograph might be able to measure the spectral type from the CO band at approximately 2.3 $\mu$m.

The conclusion that the companion is K-type is mostly consistent with previous work.  As mentioned before, Judge et al. concluded that a main sequence companion would have to be a star later than G-type.  A K-dwarf has a mass of between 0.6-0.8 \msun; taking values of 1.1 - 1.2\msun \ for \del And A gives 0.3-0.4 as the mass fraction, reasonably consistent with the value of \citet{2002A&A...391..647G} of 0.5 $\pm$ 0.1.  {However, there are some discrepancies.  If the mass of delta Andromeda is so low, the age must be much larger than the reported value of 3.2 Gyr from \citet{2003ApJ...598..636D}, as sufficient time would not have passed for the star to evolve off the main sequence.  This is not totally inconsistent, as \citet{2003ApJ...598..636D} express very low confidence in the accuracy of the age measurement for this particular star.  In light of this, we suggest that the age needs to be revised significantly upwards, to over 6 Gyrs.  Another more remote possibility consistent with our data is that the companion is actually a multiple itself, such as two even lower mass dwarfs.}

The results presented here demonstrate the potential of high contrast imaging applied to medium to long-period spectroscopic binaries.  In particular, the contrast differences between main sequence stars in binaries are readily accessible to a coronagraphic system, and the information gained can improve orbit characterization, or as in our case, distinguish quickly between different companion possibilities.

\section{Acknowledgements}
We are pleased to acknowledge the Palomar Observatory staff for their excellent support, particularly Steve Kunsman.  We greatly benefited from the expert  assistance of Rick Burruss (JPL) with the adaptive optics system.  We thank the referee for a useful review, particularly for pointing out some discrepancies in our analysis and previous work, which improved the paper.  Part of this work was carried out at the Jet Propulsion Laboratory, California Institute of Technology, under contract with the National Aeronautics and Space Administration (NASA).  Michael Bottom is supported by a NASA Space Technology Research Fellowship, grant NNX13AN42H.

\bibliographystyle{apj}

\end{document}